\documentclass[prl,aps,amsfonts,amssymb,nofootinbib,twocolumn,amsmath,superscriptaddress]{revtex4-2}

\usepackage{graphicx}
\usepackage{bm}
\usepackage{hyperref}
\usepackage{float}
\usepackage{siunitx}
\usepackage{xcolor}
\usepackage{natbib}

\begin{document}

\title{Terahertz range polarization rotation in the candidate time-reversal symmetry breaking superconductor BiNi}

\author{Ralph Romero III} \email{rromero@jhu.edu}
\affiliation{William H. Miller III Department of Physics and Astronomy, The Johns Hopkins University, Baltimore, Maryland 21218, USA}

\author{Zhenisbek Tagay}
\affiliation{William H. Miller III Department of Physics and Astronomy, The Johns Hopkins University, Baltimore, Maryland 21218, USA}

\author{Jiahao Liang}
\affiliation{William H. Miller III Department of Physics and Astronomy, The Johns Hopkins University, Baltimore, Maryland 21218, USA}

\author{Jason Y. Yan}
\affiliation{William H. Miller III Department of Physics and Astronomy, The Johns Hopkins University, Baltimore, Maryland 21218, USA}

\author{Di Yue}
\affiliation{William H. Miller III Department of Physics and Astronomy, The Johns Hopkins University, Baltimore, Maryland 21218, USA}

\author{N. P. Armitage}
\affiliation{William H. Miller III Department of Physics and Astronomy, The Johns Hopkins University, Baltimore, Maryland 21218, USA}

\begin{abstract}

 Here we report the observation of time-reversal symmetry (TRS) breaking superconductivity in a BiNi bilayer using terahertz (THz) polarimetry. Leveraging a novel high-precision THz polarimetry technique, we detect, in the superconducting state and at zero magnetic field, the smallest polarization rotation of THz light measured to date. By using the MgO substrate itself as an optical resonator, we can reference the Faraday and Kerr rotations to each other.  We observe a low-frequency Kerr rotation on the order of several hundred microradians in the superconducting phase, a clear signature consistent with TRS-breaking superconductivity.  Our measurements enable direct access to the THz-range Hall conductivity. Through a Kramers-Kronig analysis, we link these low-energy measurements to prior high-frequency magneto-optic Kerr effect (MOKE) data. This connection provides critical insight into the nature of the TRS-breaking state, supporting a multiband superconducting scenario over a disordered single-band interpretation for the origin of the Kerr effect.
\end{abstract}

\maketitle


Chiral superconductors~\cite{Kallin_2016} that break time-reversal symmetry (TRS), have been the subject of intense current study as, aside from fundamental interest, they have been proposed to host Majorana zero modes, a promising platform for topological quantum computation~\cite{kitaev2003fault,Sarma_2015}. Sr$_2$RuO$_4$ was a leading candidate to host such chiral superconductivity~\cite{Sato_2017}, however recent results are incompatible with a $p_x \pm ip_y$ type order parameter leaving the form of its superconducting order parameter an open question~\cite{Mackenzie2017,Yaguchi2002,Mao2000,Taniguchi2015,Pustogow_2019}. Moreover, its T$_c$ is fairly low, $\sim$ 1.5 K, making studies and application challenging. In 2015, epitaxially grown BiNi bilayers were shown to superconduct at 4 K~\cite{Gong_2015}. This was surprising as neither of the constituent materials exhibit superconductivity by themselves at these temperatures or pressures.  It was shown shortly thereafter using a Sagnac interferometer to measure the magneto-optical Kerr effect (MOKE) that the superconducting state of the BiNi bilayer broke TRS.  It was claimed to be an even parity, $d_{xy} \pm d_{x^2-y^2}$, order parameter, due to the fact that superconductivity was believed to be occurring at the Bi surface~\cite{GongBiNi}. This assignment was later disputed  with the observation that SC appeared to occur throughout the bulk of the film~\cite{PrashantBiNi,Wang2023}. Much like Sr$_2$RuO$_4$, consensus on the symmetry of the order parameter in BiNi remains lacking to this day. 

Two leading methods of directly detecting TRS breaking superconductivity are $\mu$SR and MOKE~\cite{wysokiński2019,  Ghosh_2020}. These are powerful probes, with MOKE via Sagnac interferometry possessing a remarkable sensitivity on the order of tens of nanoradians. However, both are limited in the sense they are single frequency and spatially local, making them difficult to cross check with theory and possibly misinterpretable as arriving from magnetic impurities. Furthermore,  the energy of the optical or infrared photons ($\sim$1 eV) used in MOKE are far from typical superconducting gap energies ($\sim$1 meV or less in many discussed systems). The mismatch of the photon energy in conventional MOKE to the natural energy scale of superconductivity inherently requires high sensitivity as the expectation is that when superconductivity onsets the majority of the spectral weight rearrangement will occur at much lower optical frequencies. With this in mind, it would be beneficial to have a probe of TRS breaking with lower photon energies in order to get spectroscopic information on the most relevant energy scales.

In this Letter, we report the detection of TRS breaking SC in a BiNi bilayer via THz range polarimetry. By using a novel scheme for high precision THz polarimetry we observe in the superconducting state of BiNi at zero magnetic field the smallest rotation of the polarization of THz light to date.  By using the MgO substrate as an optical resonator, we find low frequency Kerr rotations on order of a few hundred $\mu$radians in the superconducting state.  This is consistent with TRS breaking superconductivity.   Our
experiment enables direct measurement of the THz-range Hall conductivity.  Through a Kramers-Kronig analysis we can connect the Hall effect (measured via previous MOKE) measured at eV  frequencies to this data, which is measured much closer to the gap energy scale. With this perspective we show signatures of the TRS breaking state in BiNi is more likely to originate from a multiband scenario than a dirty single band case.

\begin{figure}[]
	\includegraphics[]{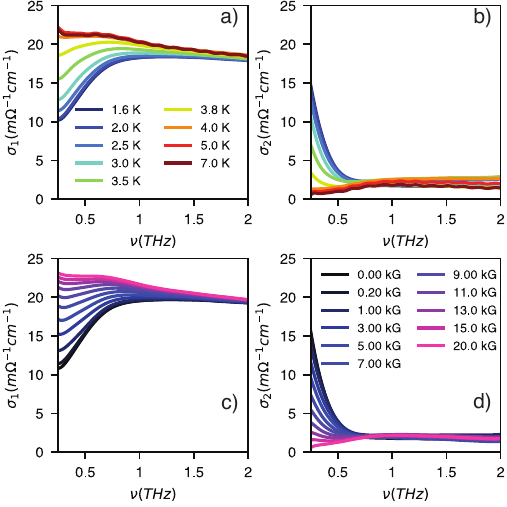}
	\caption{a),b) Temperature dependence of the real and imaginary parts of the conductivity measured using TDTS. A clear superconducting transition can be seen near 3.8 K. c), d) Magnetic field dependence of the complex conductivity. Complete suppression of superconductivity can be seen around 20 kG. This upper critical field is consistent with superconductivity coming from the Bi/Ni bilayer as opposed to NiBi$_3$.}
	\label{fig:fig1}
\end{figure}

A 10 nm thick rhombohedral Bi(110) layer was epitaxially grown on a 1 nm Ni(100) layer at 110 K, which is seeded at 300  K on a 0.5 mm thick MgO(100) substrate.  Conventional time-domain THz spectroscopy (TDTS) was performed on this sample to measure the longitudinal complex conductivity, $\tilde{\sigma}(\nu)$.  By measuring $\tilde{\sigma}(\nu)$ in the THz range while varying temperature or out-of-plane magnetic field, one can clearly identify the superconducting transition temperature, $T_c$, and the upper critical field at which SC is destroyed, $H_{c2}$.  Figs.~\ref{fig:fig1} a) and b) show the real, $\sigma_{1}(\nu)$, and imaginary, $\sigma_{2}(\nu)$, parts of the conductivity from 0.2 - 2.0 THz at temperatures from 7 to 1.6 K.  At 7 K, $\sigma_{1}(\nu)$ exhibits a slightly decreasing frequency dependence while $\sigma_{2}(\nu)$ is slightly increasing with frequency. This is indicative of a dirty metal Drude-like response with a scattering rate somewhat larger than our frequency window. As discussed in previous work, the normal state is a disordered metal whose electrodynamic response looks like neither Bi or Ni alone~\cite{PrashantBiNi}. As temperature is lowered, the onset of superconductivity is signaled by spectral weight in $\sigma_{1}(\nu)$ being pushed to a delta function at zero frequency and a corresponding 1/$\nu$ behavior developing in $\sigma_{2}(\nu)$~\cite{tinkham}. $T_c$ can thus be seen to be around 3.8 - 3.9 K. This is consistent with Ref.~\cite{PrashantBiNi} who measures $T_c$ = 4.15 K and $\Delta(1.6 K)$ = 0.67 meV. 

Figs.~\ref{fig:fig1}c) and d) show the real and imaginary conductivities at 1.6 K as we vary an external out-of-plane magnetic field. At zero field, the optical signatures of superconductivity are clear. As the field is ramped up, superconductivity is slowly suppressed and the frequency dependence begins to resemble that of higher temperatures, with superconductivity being completely killed at 20 kG. An $H_{c2}$ of this scale is consistent with the optical response being dominated by the BiNi bilayer and not NiBi$_3$ which could have been formed inadvertently at the interface~\cite{NiBi3noTRSB,Wang2023}. 

\begin{figure}[h]
	\includegraphics[]{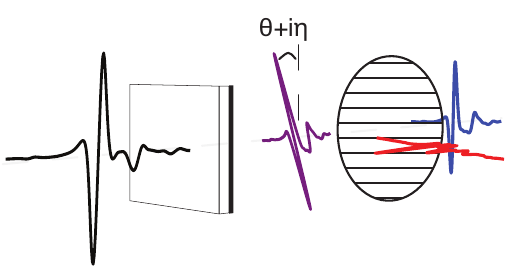}
	\caption{The operational principle of our polarization resolved time-domain THz spectrometer. A WGP is placed at \ang{45} relative to the propagation direction of the THz pulse, its wires oriented at \ang{90} relative the initial pulse polarization. This makes the WGP a polarizing beam splitter, passing the vertical component and reflecting the horizontal component of the elliptically polarized, transmitted THz pulse.   The scheme and its calibration is discussed extensively in Ref.~\cite{Zhenis}. }
		\label{fig:fig2}
\end{figure}

Next we investigate the possibility of TRS breaking in the superconducting state using a novel scheme for high-precision THz polarimetry  ~\cite{Zhenis}. Upon being transmitted through (reflected from) a sample which breaks TRS, the initially linearly polarized THz pulse is expected to emerge elliptically polarized and rotated due to the magneto optical Faraday (Kerr) effect. We use a wire grid polarizer (WGP) as a polarizing beam splitter after the sample in order to resolve the orthogonal components of the transmitted pulse. One polarization is transmitted through the WGP and one is reflected, each towards a separate detector. In this fashion $\tilde{E}_x(\nu)$ and $\tilde{E}_y(\nu)$ are measured simultaneously and the complex rotation can be extracted via the relation, $\theta_{K,1} + i \theta_{K,2} = \frac{\tilde{E}_y(\nu)}{\tilde{E}_x(\nu)}$, with a precision of better than 20 $\mu$rad ($\approx$1 millidegree)~\cite{Zhenis}.

\begin{figure}[h]
	\includegraphics[]{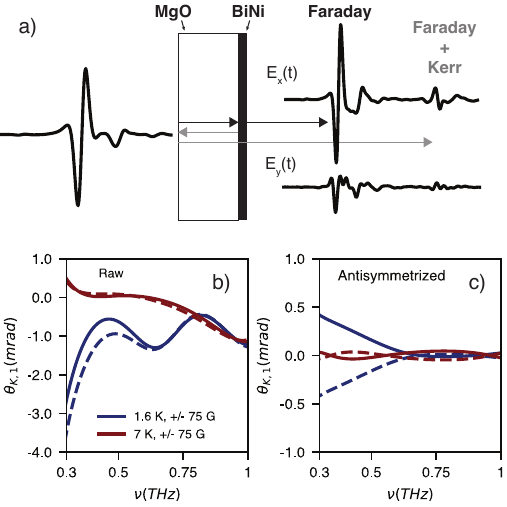}
	\caption{a) Our referencing scheme b) Raw (before antisymmetrization) real part of the  Kerr rotation, $\theta_{K,1}$, from 0.3 to 1.0 THz for the $B_{train, 1} =  + 75 G$ (solid) and  $B_{train, 2} =  - 75 G$ (dashed) at 1.6 and 7 K. b) Antisymmetrized $\theta_{K,1}$ for the same experiment. A clear inverse dependence on frequency can be seen in  $\theta_{K,1}$ at 1.6 K, when the sample is in the superconducting state.}
	\label{fig:fig3}\
\end{figure}

\begin{figure*}[]
	\includegraphics[]{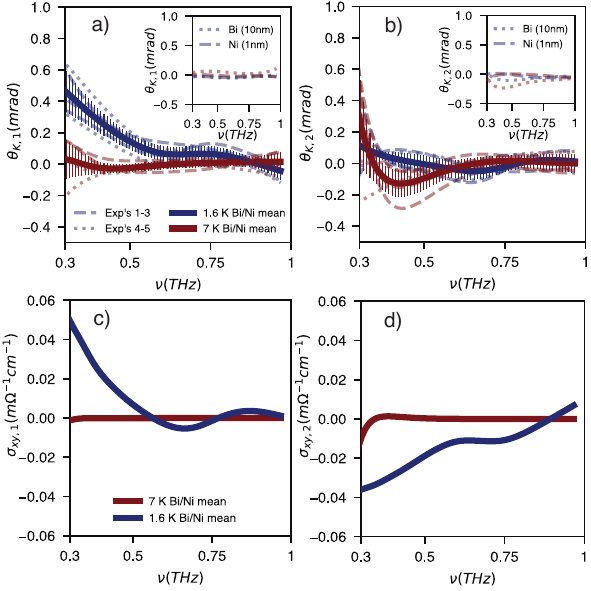}
	\caption{a) Real part of the Kerr rotation, $\theta_{K,1}$, from 0.3 to 1 THz at 1.6 and 7 K for the five separate experiments in Table 1. The bold curves are the averages of the five experiments with the error bars being one standard deviation. b) Imaginary part of the Kerr rotation, $\theta_{K,2}$, for the  separate experiments. The insets show the same quantities for $B_{train, 1/2} =  \pm 75 G$ on bare Bi (10 nm) and bare Ni (1nm). No onset of $\theta_{K,1}$ can be seen at 1.6 K in either film. c) Average of the real part of the Hall conductivity of the five experiments computed from Eq (\ref{thetaKTHz}) using $\tilde\sigma_{xx}$ in superconducting and normal state from Fig \ref{fig:fig1} and $\tilde\theta_K$ from above. d) Imaginary part of the Hall conductivity computed from Eq (\ref{thetaKTHz})}
	\label{fig:fig4}
\end{figure*}

In order to achieve such precision,  the two detectors must be calibrated such that any difference in path length and spectral response are accounted for,  a reference electric field must be defined, and the measured electric fields must be antisymmetrized with respect to magnetic field. Typically the reference is taken to be the pulse transmitted through a bare substrate,  however moving the sample stage vertically introduces considerable irreproducibility when the signal is on the order of the precision of the spectrometer. In order to meet this challenge, we have used the MgO substrate as an optical resonator to measure the Kerr rotation of the first reflected THz pulse using the initial transmitted pulse's Faraday rotation as the reference (see Fig 3a). As the pulse is first incident on the metallic thin film, the transmitted part incurs a Faraday rotation as it passes through the TRS breaking film. We use this as our reference. A part of this initial pulse which is reflected by the sample feels a Kerr rotation, is reflected off the back of the substrate and transmitted through the sample picking up a Faraday rotation similar to the first pulse. Thus by subtracting the rotation of the first pulse from the rotation of the second pulse we are left behind with a measure of the complex Kerr rotation, $\tilde{\theta}_K^{film}(\nu)$.

We use this high precision THz polarimetry to characterize TRS breaking in this material.  As is customary in the diagnosis of TRS breaking states of matter, a small magnetic training field is employed to ensure a monodomain sample.  A small external magnetic field ($B_{train} < B_{c1}$) is applied as the sample is cooled through T$_c$.   We ensured that no large remnant fields were present in the magnet by first warming the magnet above its critical temperature before every cool down.  The field is set to the training field and the sample cooled.  Once at the lowest temperature the external field is set to zero and the THz response is measured in the superconducting state (1.6K) and normal state (7K) on warm up.   As shown in Fig.~\ref{fig:fig3} b), the raw rotation data shows appreciable difference between the normal state and base temperature, but only a small difference between positive and negative training fields.   We attribute the large temperature dependent difference to a background ``rotation" that comes from (still small) misalignments and birefringence of optical elements in the beam path.

In order to largely remove the effects of misalignments and birefringence of optical elements (e.g. mirrors, cryostat windows) before and after the sample (which as discussed in Ref.~\cite{Zhenis} present problems in referencing as the matrices representing them do not generally commute) the measured electric fields are antisymmetrized with respect to the applied magnetic field, i.e. $\tilde{E}(\textbf{B}) = \frac{1}{2}(\tilde{E}(\textbf{B}) - \tilde{E}(-\textbf{B}))$.  In order to antisymmetrize the data, each experiment consists of two separate cooldowns, one with $B_{train, 1} \gtrless 0$ and the other with $B_{train, 2}  \lessgtr 0$.  At 1.6 K, the raw rotations for  $B_{train, 1} =  + 75 G$ and $B_{train, 2} =  - 75 G$ (Fig.~\ref{fig:fig3} b) show a clear training field dependence from 0.3-0.6 THz, whereas at 7 K the rotations from the two separate cooldowns are indistinguishable.  As shown in Fig.~\ref{fig:fig3} c), after antisymmetrization a decreasing frequency dependence of $\theta_{K,1}(\nu)$ is evident in the superconducting state and the rotation in the normal state is negligible.

\begin{table}[]
\caption{Training Fields} 
\centering 
\begin{tabular}{c c c} 
\hline\hline 
Experiment & $B_{train, 1} (G)$ & $B_{train, 2} (G) $ \\ [0.5ex] 
\hline 
1 & +75 & -75\\ 
2 & -75  & +75 \\
3 & +75 & -75 \\
4 & +75 & -190 \\
5 & -75  & +190 \\ 
\hline 
\end{tabular}
\label{table:Btrain} 
\end{table} 

Now we discuss the relative magnitude and origin of the observed rotation.  If the observed rotation arose from trapped flux in the superconductor,  we may expect the rotation to scale with $B_{train}$ as more flux lines are injected and pinned once $B_{train}$ was removed. However, if the rotation is due to the pairing symmetry of the superconducting order parameter breaking TRS, we may expect the signal to follow the sign of the training field, but be independent of its magnitude.  In Figures~\ref{fig:fig4} a) and b) we show the real, $\theta_{K,1}(\nu)$, and imaginary parts, $\theta_{K,2}(\nu)$ of the Kerr rotation for five separate experiments at 1.6 and 7K whose training fields are shown in Table 1. To check the repeatability and trainability of the spontaneous TRS breaking signal below T$_c$ we perform experiments (1-3) where $\lvert B_{train, 1} \rvert = \lvert B_{train, 2} \rvert $. For all three experiments, the rotation in the superconducting state always follows the training field and has similar magnitude and frequency dependence regardless of the sign of $B_{train, 1}$.
To next check that the TRS breaking is not due to pinned vortices we perform two experiments (4-5)  where $\lvert B_{train, 1} \rvert \neq \lvert B_{train, 2} \rvert $. If the TRS breaking arose from trapped vortices, the unequal training fields would induce rotations with unequal magnitudes.  Moreover the overall magnitude would be significantly larger for +190G training field relative to that of +75 G. For both we see that the rotations follow the training field and have the same magnitude and frequency dependence as the $\pm$75G experiments. The corresponding imaginary part of the rotation is approximately zero at 1.6 and 7K for all experiments. Moreover, measurements performed on bare MgO/(1nm) Ni and MgO/ (10nm) Bi with $B_{train, 1/2} =  \pm 75 G$ (inset to Fig.~\ref{fig:fig4} a), b)) show no onset of rotation in the real part below T$_c$ in either the Bi or Ni films, while the imaginary part shows similar indistinguishability from zero.

At the lowest measured frequency (0.3  THz = 1.2 meV),  $\tilde\theta_{K}$ is $\sim 0.5 + i 0$ mrad, the smallest rotation measured in the THz frequency range to date. Following the derivation of~\cite{LiangScience,DirkHgTe}, one can apply the relevant Fresnel equations in order to express the response in terms of the longitudinal and transverse THz conductivities (See Supplemental Information), $\tilde\sigma_{xx}$ and $\tilde\sigma_{xy}$,

\begin{equation}
    \tilde\theta_K^{film} (\nu) \approx - \frac{2 \tilde\sigma_{xy}}{Z_0 d (\tilde\sigma_{xx})^2}
    \label{thetaKTHz}
\end{equation}

which is valid for $\tilde\sigma_{xy} \ll \tilde\sigma_{xx} $.  With this expression we can compute the the Hall conductivity, $\tilde\sigma_{xy}$, at THz frequencies with our measured Kerr rotation, conductivity from Fig.~\ref{fig:fig1} and thickness of the sample, $d$. Using these values we solve for the Hall conductivity, shown in Figures \ref{fig:fig4} c) and d). The data shows $\tilde\sigma_{xy}$ in the superconducting decreasing in magnitude as function of frequency, where $\sigma_{xy,1} > 0$ and $\sigma_{xy,2} < 0$. In the normal state no appreciable Hall conductivity is evident.  

The above analysis further corroborates the existence of TRS breaking in the superconducting state but is this scale of the Hall conductivity consistent with previous MOKE experiments~\cite{GongBiNi}? As discussed in the SM, from general Kramers-Kronig considerations one expects that the high frequency $\sigma_{xy,1}$ goes like $(\delta/\nu)^2$ down to a cutoff energy scale $\delta$ that reflects either the superconducting gap (2$\Delta$) or the scale of interband transitions ($E_g$). The expectation is then that the Hall conductivity at high frequencies goes as $\sigma_{xy,1}(h\nu \gg E_g ) \sim \sigma_{xy,1}( 0) (\frac{E_g}{h\nu})^2$.

To extract the high frequency Hall conductivity from existing Kerr data. we performed a transmission experiment on our sample over a large spectral range. $\theta_{K}^{surface}$ from light reflected off a semi-infinite sample is related to $\tilde\sigma_{xy}$ via the equation~\cite{Okamura2020}

\begin{equation}
 \tilde\theta_K^{surface}(\nu) = -\frac{\tilde\sigma_{xy}}{\tilde\sigma_{xx}\sqrt{\tilde\epsilon_{xx}}},
    \label{thetaK1550}
\end{equation}

where all quantities are complex and its been assumed again that the Hall conductivity is much smaller than the longitudinal conductivity.  From our transmission experiment (SI Figs. 3 and 4), $\sigma_{xx}(h\nu = 0.8 eV) \approx 11.3 + i 0 \ (m\Omega^{-1}cm^{-1}) $, $\epsilon_{xx}(h\nu = 0.8 eV) \approx 2 +  i25 $  and  $\tilde\theta_{K}(h\nu = 0.8 eV) = 80 +  i0$ nrad from~\cite{GongBiNi}. We can plug these values into Eq. (~\ref{thetaK1550}) and solve for $\sigma_{xy,1}$, giving us $\sigma_{xy,1}(h\nu = 0.8 eV) \approx -3.33 \times10^{-6} \ (m\Omega^{-1}cm^{-1}) $.

Finally we can use our derived expression $\frac{\sigma_{xy,1}(\nu \gg 1)}{\sigma_{xy,1}(\nu= 0)} \sim -(\frac{E_g}{h\nu})^2$ to extract an energy scale, $E_g$. Reading off the value of $\sigma_{xy,1} (h\nu\ = 1.2 \ meV \sim 0) = 0.05 \ m\Omega^{-1}cm^{-1}$ from Fig. \ref{fig:fig4}c) gives us a value of $ E_g \approx 7$ meV. This is obviously too large to correspond to the superconducting gap of BiNi ($\approx$0.7 meV), but may correspond to the inverted bandgap at the L point of Bi of $\approx$ 15  meV~\cite{BiBandstructure}. Thereby we can conclude that there is possible rough correspondence between the Kerr rotation data at 0.8 eV and our THz data at meV scales. Moreover, since our extracted $E_g$ is an order of magnitude larger than the superconducting gap, it is most likely that this TRS breaking superconducting state originates from a multiband scenario as alluded to in the second scenario proposed in the SI.   This is corroborated by the more complicated form for the real and imaginary Hall conductivities found in Fig. 4.   More explicit calculations of the Hall response are needed for possible TRS breaking superconducting order parameters BiNi bilayers.

We would like to thank V. Yakovenko for helpful conversations, and P.-C. Xu and X. Jin for preparation of the BiNi film.   Measurements at JHU were supported by the ARO MURI “Implementation of axion electrodynamics in topological films and device” W911NF2020166. The instrumentation development at JHU that made these measurements possible was supported by the Gordon and Betty Moore Foundation EPiQS Initiative Grant GBMF- 9454. 

The data that support the findings of this article are openly available \cite{Zenodo}

\bibliography{main}

\end{document}


\title{Supplemental Information: Terahertz range polarization rotation in the candidate time-reversal symmetry breaking superconductor BiNi}

\author{Ralph Romero III} \email{rromero@jhu.edu}
\affiliation{William H. Miller III Department of Physics and Astronomy, The Johns Hopkins University, Baltimore, Maryland 21218, USA}

\author{Zhenisbek Tagay}
\affiliation{William H. Miller III Department of Physics and Astronomy, The Johns Hopkins University, Baltimore, Maryland 21218, USA}

\author{Jiahao Liang}
\affiliation{William H. Miller III Department of Physics and Astronomy, The Johns Hopkins University, Baltimore, Maryland 21218, USA}

\author{Jason Y. Yan}
\affiliation{William H. Miller III Department of Physics and Astronomy, The Johns Hopkins University, Baltimore, Maryland 21218, USA}

\author{Di Yue}
\affiliation{William H. Miller III Department of Physics and Astronomy, The Johns Hopkins University, Baltimore, Maryland 21218, USA}

\author{N. P. Armitage}
\affiliation{William H. Miller III Department of Physics and Astronomy, The Johns Hopkins University, Baltimore, Maryland 21218, USA}
\maketitle
\section{THz Kerr rotation}

As discussed in the main text, we obtain precise information about polarization rotation from the sample, by using the substrate as an optical resonator.  We reference the multiples pulses that pass through the sample against each other as they each have different histories of interaction with the film, but similar histories of transmission through the whole optical system.

The expression for the Kerr rotation obtained via subtracting the rotation of the main pulse from the rotation of the first reflected pulse can be found in the supplemental information of \cite{LiangScience,DirkHgTe}. It is, 
\begin{align*}
    \tilde\theta_K = \frac{2Z_0n\tilde\sigma_{xy}d}{\tilde n^2 - [1+2Z_0\tilde\sigma_{xx}d + Z_0^2\left((\tilde\sigma_{xx}d)^2+(\tilde\sigma_{xy}d)^2\right)] }
\end{align*}

Here $\tilde n$ is the substrate index of refraction.  By making the  assumption that $\tilde\sigma_{xx} 
\gg \tilde\sigma_{xy}$ and $Z_0 \tilde\sigma_{xx} d 
\gg \tilde n$ we can arrive at equation (1) in the main text.

\begin{align*}
    \tilde\theta_K = &\frac{2Z_0n\tilde\sigma_{xy}d}{n^2 - [1+2Z_0\tilde\sigma_{xx}d + Z_0^2\left((\tilde\sigma_{xx}d)^2\right)]
    } \\
    &\approx\frac{2Z_0n\tilde\sigma_{xy}d}{n^2 - (1+Z_0\tilde\sigma_{xx}d)^2 } \\
    &\approx \frac{2Z_0n\tilde\sigma_{xy}d}{ - (Z_0\tilde\sigma_{xx}d)^2 } = -\frac{2 \tilde\sigma_{xy}}{Z_0 d (\tilde\sigma_{xx})^2}
\end{align*}


\section{Relating optical to THz Kerr rotation}
It is possible to derive a very general relation that allows one to estimate the low frequency Hall conductivity from the high frequency tail of the non-dissipative optical Hall response.  In order to have any polarization rotation at all upon entering in the TRSB state, one must have either disorder (the dirty limit) or a multi-band system.  This has the effect in either case of putting spectral weight at finite frequency that can then be modified by the occurrence of superconductivity.   Otherwise in the single band clean limit all spectral weight is in the zero frequency delta function both above and below T$_c$.

We begin with a general Kramers-Kronig (KK) expression for the Hall conductivity, $\sigma_{1, xy}$ e.g.,
\begin{equation}
    \sigma_{1, xy}(\omega) = -\frac{2}{\pi} \int_0^\infty \frac{\omega' \sigma_{2,xy}(\omega')}{\omega'^2 - \omega^2}d\omega'.
    \label{eq1}
\end{equation}

As usual, we can find the non-dissipative (dissipative) conductivity at any frequency if we know the dissipative (non-dissipative) conductivity for all frequencies.   The dissipative part of the Hall conductivity can be expressed as the difference between dissipative conductivities for right and left circularly polarized light.  $\sigma_{2,xy}$ is the dissipative response of the Hall effect and can be expressed as
\begin{equation}
    \sigma_{2, xy} = \Im\left(\frac{\sigma_r-\sigma_l}{2i}\right) = -\left(\frac{\sigma_{1r}-\sigma_{1l}}{2}\right) .
    \label{eq2}
\end{equation}

Putting  (\ref{eq2}) $\rightarrow$ (\ref{eq1}) one has, 

\begin{equation*}
    \sigma_{1, xy}  = \frac{1}{\pi}\int_0^\infty \frac{\omega'\left(\sigma_{1r}(\omega')-\sigma_{1l}(\omega')\right)}{\omega'^2 - \omega^2}d\omega'.
\end{equation*}

\begin{figure}
    \centering
    \includegraphics[]{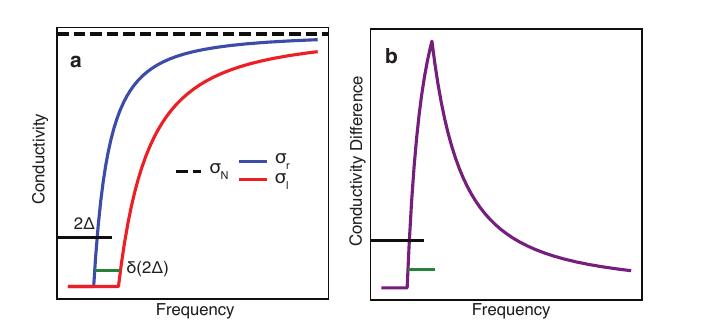}
    \caption{a) In the dirty limit case shown here, the real part of the conductivity for a TRSB SC which has a different superconducting gap for the right and left absorptive channels. The energy scales, $2\Delta$ and $\delta(2\Delta)$ are denoted in black and green respectively. b) The difference of the conductivities in a). As emphasized in the derivation when this quantity is finite it is of order the normal state conductivity $\sigma_N$, falling off quickly with increasing frequency. }
    \label{S1}
\end{figure}

Let us make the specific assumption for a time reversal symmetry breaking superconductor (TRSB SC), that system is in the dirty limit (so the dissipative spectral response is flat at frequencies of order the superconductivity gap), and there is different superconducting gap apparent in $\sigma_r$ and $\sigma_l$ the difference of which we take as $\delta (2\Delta) = 2\Delta_r- 2\Delta_l$ with an average value of $2\Delta$.  This is a specific assumption here, but for our expression to be valid it will only be necessary that the spectral weight in the dissipative Hall response be concentrated in a narrow spectral region.  It could be narrow concentrated near the superconducting gap edge as in the dirty limit first considered here.  Or it could be that interband transitions are gapped out by the occurrence of superconductivity in the multiband case considered below.  For the dirty limit case considered first, we take  $\sigma_{2, xy}$ to be non-zero only in a small energy range  $\delta (2\Delta) $ centered around $2 \Delta$ as shown in Fig. S\ref{S1} b.  Also note that we are using units where $\Delta$ is expressed in angular frequency e.g. energy divided by $\hbar$.

For the dirty limit, the region where the quantity $ \sigma_{1  r} - \sigma_{1 l}$ is nonzero it is of order the normal state conductivity $\sigma_N$.  We perform the KK transform for the high frequency limit, which is at at frequencies well above the region where $\sigma_{1  r} - \sigma_{1 l}$ is finite.  As noted above, $\sigma_{1  r} - \sigma_{1 l} $ is appreciably different from zero only in a small frequency range $\delta (2 \Delta)$ around $2\Delta$. Therefore for $\omega \gg 2\Delta$, we can write

\begin{equation}
    \sigma_{1, xy}(\omega) \approx -\frac{1}{\pi}\frac{1}{\omega^2}\int_0^\infty  \omega' (\sigma_{1  r}(\omega') - \sigma_{1 l}(\omega') ) d\omega'.
    \label{eq3}
\end{equation}

 Furthermore, because $\delta \Delta \ll \Delta$,  the $\omega'$ inside the integral can be taken to be its mean value in the region where $\Delta\sigma_{1 \ r,l} \neq  0$ and pulled outside the integral
 
 \begin{align}
    &= -\frac{1}{\pi}\frac{2\Delta }{\omega^2} \int_0^\infty ( \sigma_{1  r}(\omega') - \sigma_{1 l}(\omega'))  d\omega'.
\end{align}
 

The integral is now simply the differences of spectral weights between the right and left absorptive channels.  Putting it all together we have

 \begin{align}
  \lim_{ \omega \to  \infty } \sigma_{1, xy}(\omega)  = -\frac{1}{\pi}\frac{2\Delta }{\omega^2}  \gamma \sigma_N \delta(2  \Delta) .
      \label{HighFreq}
\end{align}
Here $\gamma$ is a factor that accounts for the modifications of the optical matrix elements for above  gap absorptions.  For the case of a dirty limit BCS superconductor $\gamma = \pi/2$.  It is zero for a clean limit single-band superconductor as there is no finite frequency absorption for such a system.   But it also does not factor in our final result.  To recap, this is an expression that is valid for frequencies well above the region of absorption ($\omega \gg 2\Delta$) if the off-diagonal absorption is found in a narrow range of frequencies (e.g. $\delta(2 \Delta ) \ll2 \Delta$).

We can derive an analogous expression for the low frequency case  $\omega \ll 2\Delta$,
\begin{align*}
    \sigma_{1, xy}(\omega) &= \frac{1}{\pi}\int_0^\infty\frac{\omega' ( \sigma_{1  r}(\omega') - \sigma_{1 l}(\omega')) }{\omega'^2(1-\frac{\omega^2}{\omega'^2})}d\omega',\\   \sigma_{1, xy} (\omega \ll \omega') 
    &\approx  \frac{1}{\pi}\int_0^\infty\frac{\sigma_{1  r}(\omega') - \sigma_{1 l}(\omega')}{\omega'}[1+\frac{\omega^2}{\omega'^2}]d\omega',\\ \sigma_{1, xy} (\omega \ll 2 \Delta )
    & \approx  \frac{1}{\pi} \frac{1}{2\Delta}  [1+\frac{\omega^2}{ (2\Delta )^2}]\int_0^\infty  (\sigma_{1  r}(\omega') - \sigma_{1 l}(\omega') ) d\omega',\\
\end{align*}

Here we have again used the fact that the integral is only non-zero in a narrow frequency around $2\Delta$ and that $\omega'$ can be replaced in the integral by its mean value and then taken out as a constant.  The integral is again the difference is absorption for right and left circularly polarized light.

\begin{align*}
   \sigma_{1, xy} (\omega) &=  \frac{1}{\pi} \frac{  \gamma \sigma_N \delta(2  \Delta)}{2\Delta}  [1+\frac{\omega^2}{ (2\Delta )^2}] ,\\
\end{align*}

Note this form is constant as  $\omega\rightarrow0$ with a small quadratic correction that is small as long as the frequencies being considered are well below the gap scale.

\begin{equation}
    \sigma_{1, xy}(0)=  \frac{1}{\pi}\frac{\gamma\sigma_N \delta(2\Delta)}{2\Delta}
    \label{LowFreq}
\end{equation}

Now we can combine Eqs. \ref{HighFreq} and  \ref{LowFreq} to get our key result.

\begin{align}
\frac{\sigma_{1, xy}(\omega)}{\sigma_{1, xy}(0)}= - \left(\frac{2\Delta}{\hbar\omega}\right)^2.
    \label{DirtyLimit}
\end{align}

This is a general and useful result that can be used to relate the low frequency Hall response to the high frequency optical response for dirty limit TRSB superconductor.   It is valid for frequencies well away from the range of optical absorptions that are different for right and left hand light and depends only on the validity of the KK relations.

\begin{figure}
    \centering
    \includegraphics[]{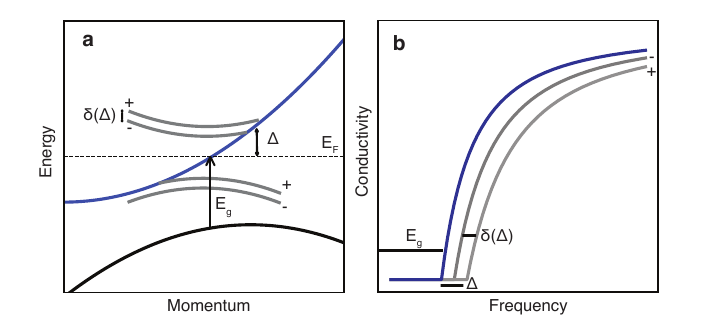}
    \caption{a) Schematic of the band structure for TRSB SC in a multiband system. The grey bands are the spin split Bogoliubon branches which modify the  optical conductivity as shown in b). b) The optical conductivity for this system. Although microscopically very different from the scenario in Fig S\ref{S1}, the key assumption from earlier still holds i.e. $\sigma_{1  +} - \sigma_{1 -} $ is nonzero in a small frequency range. }
    \label{S2}
\end{figure} 

For a superconductor in the clean limit, one must have multiple bands in order to see signatures of superconductivity in the optical response.  In a multi-band system changes to the optical spectra result because interband transition change because near-E$_F$ states are blocked because of the opening of the superconducting gap.   Therefore the frequency scale that changes occur at are not the superconducting gap scale, but the scale of the interband transition energy $E_g$.   See Fig. (S\ref{S2} a) for a schematic.  Again, in the high frequency limit e.g. at frequencies well above the range for absorptions that distinguishes right and left light e.g.   $ \omega \gg E_g $ we have,
\begin{equation}
    \sigma_{1, xy} (\omega) = -\frac{1}{\pi}\frac{E_g}{\omega^2} \gamma \sigma_N  \delta \Delta .
    \label{HighMulti}
\end{equation}

We retain the spirit of the dirty limit single band case above, but now $\sigma_N$ is the normal state conductivity at the band gap frequency and $\gamma$ is a factor that accounts for the modification of the optical matrix elements due to the onset of superconductivity.  Again, these factors do not factor in the final result.

At low frequencies we can also derive an analogous expression as we did for the dirty limit case

\begin{equation}
    \sigma_{1, xy} (0) =  \frac{1}{\pi}\frac{\gamma \sigma_n \delta \Delta }{E_g } .
      \label{LowMulti}
\end{equation}

Combining Eqs. \ref{HighMulti} and \ref{LowMulti} we have the expression

\begin{equation}
        \frac{\sigma_{1, xy}(\omega)}{\sigma_{1, xy}(0)}= - \left(\frac{E_g}{\hbar\omega}\right)^2.
        \label{MultiBand}
\end{equation}

Obviously Eqs. \ref{DirtyLimit} and \ref{MultiBand} share similar structure and for the same reason.  Both the low and high frequency Hall responses are governed by the spectral weight of the absorption {\it difference} that distinguishes right and left and the energy range that this absorption is found in.   Also note that $\gamma$ has a numeric value quite different than the BCS case, but it does not feature in the final result Eq. \ref{MultiBand}.  These expression are also clearly compatible with existing theories for TRSB superconductors~\cite{Kallin_2016,lutchyn2009frequency,brydon2019loop,can2021probing}.

\clearpage
\section{Optical Data}

\begin{figure}[h]
    \centering
    \includegraphics[width=4in]{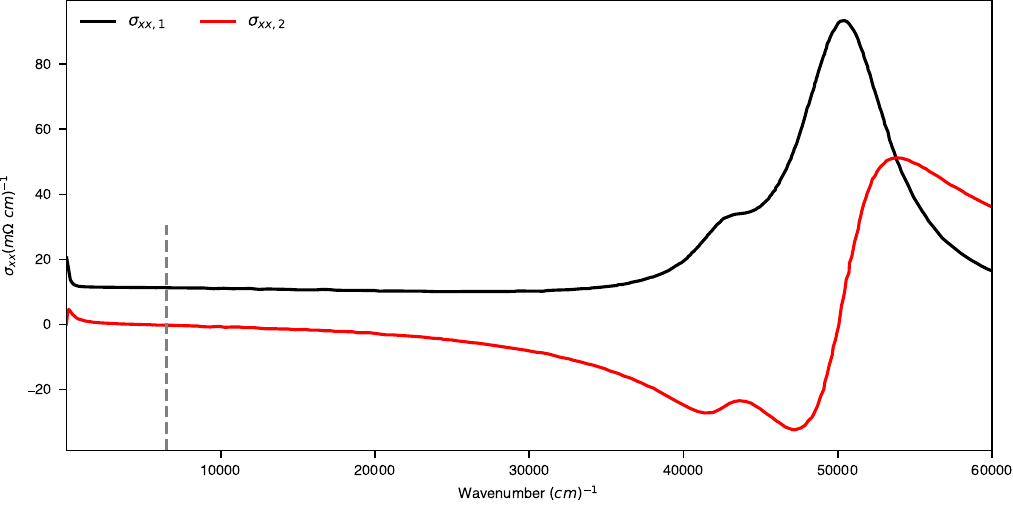}
    \caption{ The real (black) and imaginary (red) part of the longitudinal conductivity for the BiNi thin film at optical frequencies obtained with UV-Vis spectrometer and FTIR spectrometer. Recall 1550 nm = 6452 cm$^{-1}$ = 0.8 eV/h indicated by the grey dashed line.}
    \label{S3}
\end{figure}

\begin{figure}[h]
    \centering
    \includegraphics[width=4in]{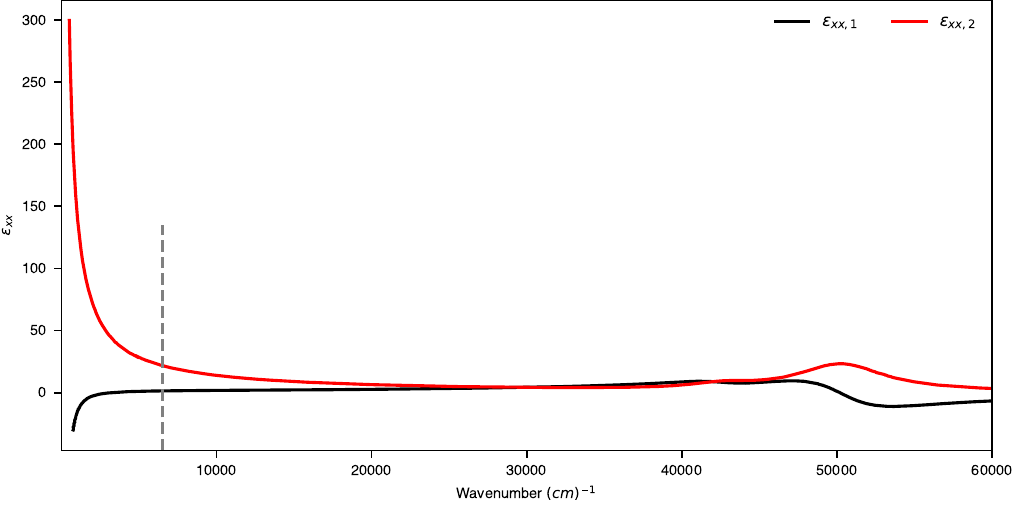}
    \caption{ The real (black) and imaginary (red) part of the dielectric function for the BiNi thin film at optical frequencies obtained with UV-Vis spectrometer and FTIR spectrometer. Recall 1550 nm = 6452 cm$^{-1}$ = 0.8 eV/h indicated by the grey dashed line. }
    \label{S4}
\end{figure}

Here we show the results of room temperature UV-Vis and Fourier Transform Infra-Red transmission (FTIR) measurements. Experiments were done in Cary 5000 UV-Vis-NIR Spectrophotometer (3000 $cm^{-1}$ to 57500 $cm^{-1}$) and Bruker V80 Spectrometer (1500 $cm^{-1}$ to 7000 $cm^{-1}$). The transmission spectrum in Mid-IR region were measured using a Globar source and mercury cadmium telluride (MCT) detector. We analyzed both the resulting transmission data and the THz conductivity with Reffit using multi-layer dielectric function model.\cite{RefFit} The resulting optical conductivity and dielectric constant data are shown in Figs. \ref{S3} or \ref{S4}.

\bibliography{main}